\documentclass[%
 reprint,
 amsmath,amssymb,
 aps,
]{revtex4-2}

\usepackage{graphicx}
\usepackage{dcolumn}
\usepackage{bm}

\begin{document}

\preprint{APS/123-QED}

\title{First results from a search for \\ coherent elastic neutrino-nucleus scattering (CE$\nu$NS) at a reactor site}

\author{J. Colaresi$^{1}$}
\author{J.I. Collar$^{2}$}
\email{collar@uchicago.edu}
\author{T.W. Hossbach$^{3}$}
\author{A.R.L. Kavner$^{2}$}
\author{C.M. Lewis$^{2}$}
\author{A.E. Robinson$^{4}$}
\author{K.M. Yocum$^{1}$}
\affiliation{%
$^{1}$Mirion Technologies Canberra, 800 Research Parkway, Meriden, CT, 06450, USA
}%
\affiliation{%
$^{2}$Enrico Fermi Institute,
University of Chicago, Chicago, Illinois 60637, USA
}%
\affiliation{%
$^{3}$Pacific Northwest National Laboratory, Richland, Washington 99354, USA
}%
\affiliation{%
$^{4}$D\'{e}partement de Physique, Universit\'{e} de Montr\'{e}al, Montr\'{e}al, H3C 3J7, Canada
}%


\date{\today}

\begin{abstract}
The deployment of a low-noise 3 kg p-type point contact germanium detector at the \mbox{Dresden-II} power reactor, 8 meters from its 2.96 GW$_{th}$ core, is described. This location provides an unprecedented (anti)neutrino flux of 8.1$\times 10^{13} ~\bar{\nu_{e}}/$cm$^{2}$s. When combined with the 0.2 keV$_{ee}$ detector threshold achieved, a first measurement of CE$\nu$NS from a reactor source appears to be within reach. We report on the characterization and abatement of backgrounds during initial runs, deriving improved limits on extensions of the Standard Model involving a light vector mediator, from preliminary data.
\end{abstract}

\maketitle



Neutrinos with energy below few tens of MeV can scatter coherently from the nucleus as a whole via the weak  neutral current, producing in the process a low-energy nuclear recoil (NR). This mechanism was first described in 1974 \cite{freedman}, and observed  in 2017 using spallation neutrinos scattering off a dedicated low-background CsI[Na] scintillator \cite{science,ournim,bjorn,nicolethesis}. This uncontroversial Standard Model (SM) prediction is typically referred to as Coherent Elastic Neutrino-Nucleus Scattering (CE$\nu$NS). The large enhancement to the scattering cross section brought about by the coherent nature of the interaction results in a neutrino detector miniaturization with potential technological applications. CE$\nu$NS experimentation also provides numerous new opportunities to search for  physics beyond the SM, and for the study of nuclear structure. 

Experiments at upcoming intense spallation sources will continue to expand the physics reach of CE$\nu$NS \cite{ESS,ivan,dimitrios2}. However, the much higher flux of  electron (anti)neutrinos in the vicinity of power nuclear reactors generates an alternative suitable source for CE$\nu$NS studies, providing an enhanced signal rate, and a complementary sensitivity in most areas of phenomenological interest reachable through this new coupling \cite{dimitrios,manfred}. Reactor sources are nevertheless expected to produce NRs considerably lower in energy (sub-keV) than those from spallation sources (few keV). The benefits to background reduction from the use of a pulsed spallation source \cite{science} are also absent. This generates considerable difficulties in the selection of a viable detector technology for this alternative approach to CE$\nu$NS measurements. 

\begin{figure}[!htbp]
\includegraphics[width=.87 \linewidth]{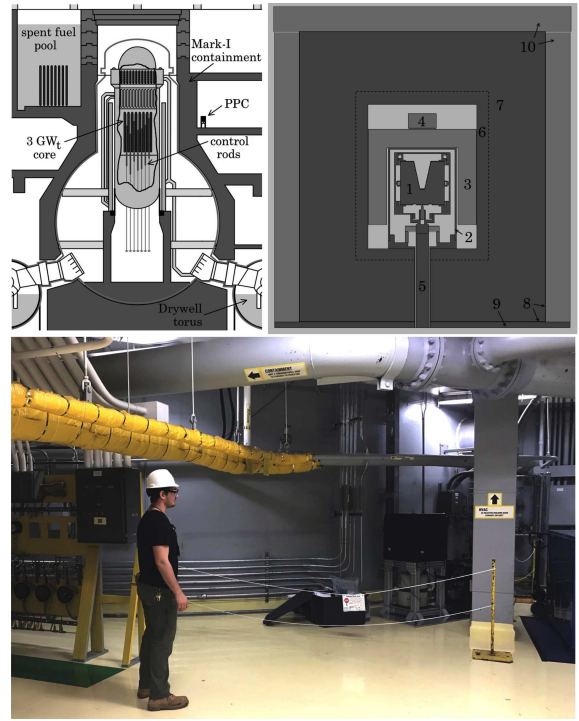}
\caption{\label{fig:epsart} {\it Top left:}  PPC detector location within the Mark-I design of the Dresden-II boiling water reactor (BWR) \cite{wiki}. {\it Top right:} cross section of the detector and shield (see text). {\it Bottom:} Compact footprint of the assembly, next to the cylindrical primary containment wall. A small cart containing all electronics is visible behind  the column, next to the detector. }
\end{figure}

P-type point contact (PPC) germanium detectors have been proposed as a technology up to this challenge \cite{PPC}. They provide a unique combination of sufficiently-large target mass, sub-keV energy threshold, and excellent intrinsic radiopurity, while bypassing the limited charge-collection efficiency and degraded energy resolution characteristic of n-type point contact alternatives \cite{Luke}. At low energy, PPCs allow to discriminate surface events from those taking place in the bulk of the germanium crystal \cite{surface}. At higher energies, they provide the ability to differentiate single-site from multiple-site interactions \cite{PPC}. As a result, they have found  applications in dark matter searches \cite{cogent,cdex}, neutrinoless double-beta decay experiments \cite{majorana,gerda,legend}, ongoing attempts at CE$\nu$NS detection \cite{ESS,conus1,texono,nugen}, and exotic decays \cite{muon}. A  compact detector profile, illustrated by this work, allows to envision their use as eminently-fieldable reactor monitoring devices with nuclear non-proliferation applications \cite{adam,huber}.

PPCs of  inverted coaxial (IC) design \cite{david} provide optimal charge collection in a multi-kg germanium crystal. In this work we employ the largest (2.924 kg) PPC in operation, one  of this style. The device is dubbed NCC-1701, in reference to ``Neutrino Coherent Coupling", and to the detector serial number assigned by the manufacturer, Mirion Technologies. The advantages of an IC PPC for the present application are multiple: first, the large peak-to-Compton ratio of a crystal this massive reduces low-energy backgrounds. A coadjutant effect in this same respect is the smaller fraction of the crystal mass represented by the thin (sub-mm) transition layer responsible for slow rise-time, low-energy surface events from incomplete charge collection \cite{surface}. Second, the distances travelled to the electrodes by charge carriers  are minimized in this design, resulting in faster rise-times for CE$\nu$NS events uniformly distributed  in the bulk of the crystal, facilitating their identification. Finally, the concentration of a large target mass into a single cryostat results in a compact radiation shield, ideal for reactor monitoring applications. The NCC-1701 assembly has a footprint of just 60 cm $\times$ 60 cm. It was installed within a single day, by three workers (Fig.\ 1). 

In anticipation to the use of an electric cryocooler \cite{cryocycle}, the internal design of the detector was revised so as to reduce a parasitic capacitance able to translate small vibrations into low-energy microphonic events \cite{vibration1,vibration2}. During laboratory tests, no correlation between detector noise and cryocooler status (on/off) could be observed. This modification  also proved to be  advantageous  in the industrial environment of the reactor site, rich in acoustic noise and mechanical vibrations. Prior to deployment, the commercial preamplifier of the detector was altered to increase its gain by $\times$12. This rendered the  noise of the digitizer employed for data-acquisition (DAQ)  negligible in comparison to the intrinsic detector noise. The temperature and settings of the field-effect transistor responsible for the first stage of signal amplification  were optimized for noise reduction, and the decay constant of the preamplifier output was elongated. The latter  allows to profit from longer integration time constants during shaping of the preamplifier signal, improving energy resolution. The cumulative effect of these measures yielded a pulser noise figure of 91 eV FWHM, at a laboratory temperature of 20$^{\circ}$C. Next-generation multi-kg PPCs are expected to reach a noise level below 50 eV FWHM \cite{pascal}. 

At the reactor site, an elevated ambient temperature approaching 35$^{\circ}$C during summer  was observed to increase  pulser noise  to 154 eV FWHM for presently discussed runs. This effect is expected from the dependence of detector leakage current, which drives the parallel component of noise, on crystal temperature \cite{leakage}. Impact on energy resolution was minimized through the use of a 36 $\mu$s zero-area cusp filter \cite{zac1,zac2,zac3} during off-line pulse shaping. An improvement in this respect can be expected from alternative cryocooler systems able to dynamically respond to ambient temperature changes \cite{newcryo}, or active temperature control in the detector vicinity.

\begin{figure}[!htbp]
\includegraphics[width=.9 \linewidth]{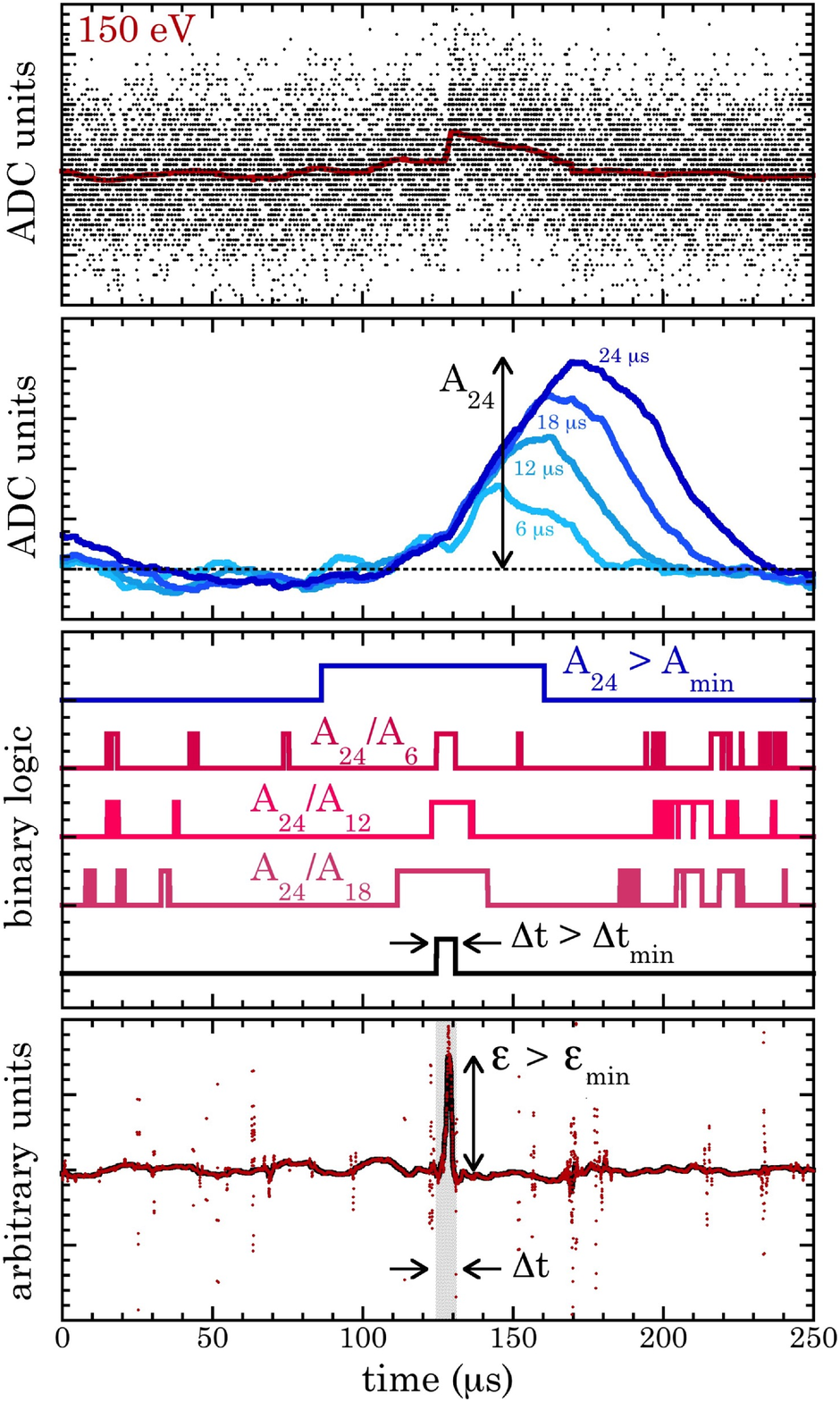}
\caption{\label{fig:epsart} Steps in data filtering, illustrated for a 150 eV$_{ee}$ signal in a 78 eV FWHM point-contact detector. From top to bottom: 1)  preamplifier waveform digitized at 120 MS/s. A red line shows the wavelet-denoised trace, obtained offline, highlighting the characteristic rise-time and longer decay-time of a radiation-induced event.\ 2) FPGA trapezoidal shaping of the waveform, using four integration constants.\ 3) Real-time  logic-level  conditions described in the text, offset by the peaking time for the $t$ = 24 $\mu$s filter. The bottom trace shows a trigger-generating coincidence among them.\ 4) Offline edge-finding. Dots show the fast derivative of the denoised trace in 1), a black line joining them following median-filtering.   }
\end{figure}

Fig.\ 1 displays a labelled cross-section of the NCC-1701 detector and shielding: 1) PPC crystal, 2) electroformed OFHC copper cryostat endcap, 3) inner plastic scintillator veto, 4) Hamamatsu R6041 photomultiplier (PMT), 5) cryostat coldfinger, 6) 2.5 cm-thick low-background lead layer, 7) 12.5 cm-thick regular lead layer, 8) 0.6 mm-thick cadmium sheet (4$\pi$ coverage), 9) steel table, 10) 5 cm-thick plastic scintillator outer veto with built-in PMTs (five-side coverage). Attention was paid to the radiopurity and cleaning of  internal PPC components and those in the inner veto, achieving a background level of 25 counts/keV-kg-day at 0.2 keV$_{ee}$ (``ee" stands for ``electron equivalent", i.e. ionization energy), under a 6 meter of water equivalent laboratory overburden \cite{ESS}. 

The main purpose of the inner veto is to reject fast neutrons able to contribute to the CE$\nu$NS energy region of interest (ROI) via elastic scattering \cite{ESS}. Its small, low-background photomultiplier can be operated at single-photoelectron (PE) sensitivity without a significant dead-time penalty. Its light-collection efficiency (8.5\% minimum) was measured and simulated over its volume, making use of a recently-validated model of organic scintillator response to low-energy proton recoils \cite{ej301}, demonstrating its ability to tag events produced by neutrons of energy above few tens of keV. A second  use for this inner veto is to reinforce the efficiency of the external veto in tagging cosmic ray-induced events, of importance in a reactor site without   significant overburden (Fig.\ 1). As a present precaution, the inner veto was operated at a reduced sensitivity corresponding to a 3 PE threshold, to avoid  dead-time episodes introduced by transient PMT ringing, noticed during laboratory tests. 

Of particular interest is the use of a real-time triggering algorithm as part of the DAQ, implemented through an AC-coupled four-channel fast digitizer (NI 5734)
embedded in a field-programmable gate array (FPGA) platform (NI PXIe-7966R). It imposes a noise-filtering  condition previously exploited via analog electronics in searches for exotic processes \cite{goulding1,goulding2,cogent,cdex,gemma}, capable of rejecting low-energy events produced by microphonics and other disturbances to the preamplifier output trace. It relies on the observation that the ratio of pulse amplitudes obtained with dissimilar shaping-filter integration times is an energy-independent constant for radiation-induced, well-formed preamplifier signals. In contrast to this, irregular preamplifier signals from microphonics and other low-energy nuisances, deviate from this constant ratio. The FPGA is programmed to shape the streaming digitized preamplifier output using four trapezoidal filters, yielding shaped amplitudes $A_{t}$, where $t$ is the shaping time in $\mu$s (Fig.\ 2). The FPGA continuously inspects the three unique ratios between these four amplitudes, looking for instances of their agreement with pre-determined ranges of accepted values, established using electronic pulser and radiation source calibrations \cite{PPC}. The FPGA triggers, fetching and saving  waveforms, only when these three conditions are simultaneously met over a time interval longer than a user-defined $\Delta t_{min}$ (Fig.\ 2), in coincidence with an amplitude threshold $A_{24} > A_{min}$  also being surpassed (of the four filters, $t=24  ~\mu$s provides the lowest detector noise). $A_{min}$ can be adjusted to provide a suitable trigger rate and signal acceptance. The DAQ monitors a high-gain channel as described, but saved waveforms include a low-gain channel recording signals of up to 900 keV$_{ee}$, and a third channel for  logic signals from the vetoes. 

A separate ``edge-finding" condition is imposed during offline analysis: the median-filtered fast derivative of the wavelet-denoised preamplifier trace is inspected, looking for the presence of a threshold condition $\varepsilon>\varepsilon_{min}$ within the FPGA trigger window  (Fig.\ 2). This powerful additional test confirms the presence of the rising edge characteristic of radiation-induced pulses. It reduces the energy threshold of the detector, by discarding the small fraction of low-frequency ripples in the preamplifier output that still manage to confound the FPGA logic. The time window $\Delta t_{min}$ is made large enough (few $\mu$s) to accept both bulk and surface events. Additional ``quality cuts" can be imposed based on the requirement that a minimum span of time be sustained above  $\varepsilon_{min}$, and on the location of the trigger within the fetched waveform trace. These are particularly indicated for a complete removal of temperature-correlated events near threshold, caused by cryocooler vibrations when operating at high power. 

This  FPGA ``intelligent trigger" algorithm was of crucial importance to accomplish a stable 0.2 keV$_{ee}$ threshold, while keeping the rate of data-writing to disk manageable at $\sim$20 Hz. In future work we plan to migrate the edge-finding condition to the FPGA. This will allow to further reduce $A_{min}$, improving energy threshold and signal acceptance, while preserving the same modest data throughput. Compact FPGA-based DAQ systems (Fig.\ 1) should be considered an intrinsic part of any realistic future application of PPCs to reactor monitoring.

\begin{figure}[!htbp]
\includegraphics[width=.9 \linewidth]{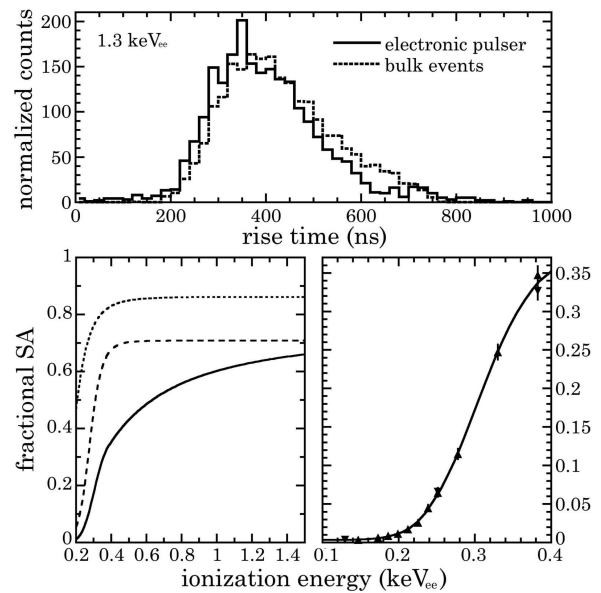}
\caption{\label{fig:epsart} {\it Top:} similar rise-times for 1.3 keV$_{ee}$ electronic pulser signals and those from $^{71}$Ge L-shell EC in the bulk of the PPC, displaying a characteristic lognormal distribution \cite{cogent}. {\it Bottom left:} pulser signal acceptance following  progressive cuts. Dotted curve: FPGA trigger efficiency and offline edge-finding condition. Dashed: addition of quality cuts. Solid: addition of rise-time cut. {\it Bottom right:} detail of the solid curve, showing discrete pulser measurements. Error bars are statistical, often encumbered by datapoints. Different symbols are used for two pulser runs separated by a few days. }
\end{figure}

Prior to reactor-site installation,  information on the radioactive environment at the proposed detector emplacement was limited to knowledge of a $^{60}$Co contamination in a nearby pipe \cite{co60}, producing a gamma equivalent dose of 1.5 mrem/hr in the absence of shielding. This was deemed to be as low as can realistically be expected from a location this close (8 m) to the axial midpoint of a BWR fuel assembly. Laboratory tests were performed using intense $^{22}$Na and $^{88}$Y gamma sources positioned to produce this same dose at the PPC. The 15 cm of lead  in the shielding (Fig.\ 1) reduced their contributions to the CE$\nu$NS sub-keV  ROI to a level more than an order of magnitude below the best background later obtained for reactor operation at full 2.96 GW$_{th}$ power (Rx-ON). Additionally, as expected from the peak-to-Compton ratio of this large PPC, the vast majority of gamma-induced events in the sub-keV ROI appeared as surface events, rejectable via rise-time analysis.

\begin{figure}[!htbp]
\includegraphics[width=.9 \linewidth]{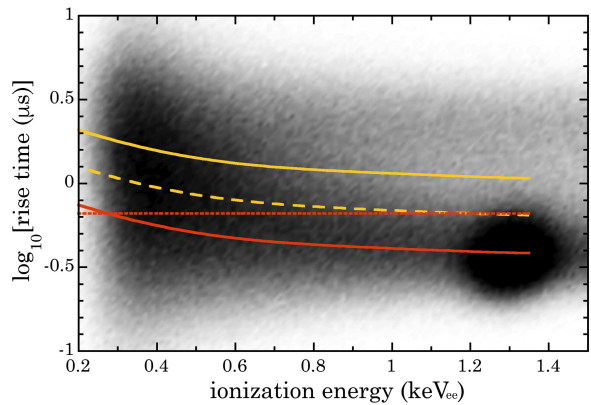}
\caption{\label{fig:epsart} Grayscale plot of ROI events passing quality cuts, for the 37 d Rx-ON run. Solid curves mark the median of rise-time distributions for surface (yellow) and  bulk (orange) events. These derive from fits to narrow (50 eV wide) energy bands, using characteristic lognormal distributions \cite{cogent}. The  median for bulk events closely follows  one obtained for electronic pulser fast signals (Fig.\ 3). The migration towards longer rise times is the effect of noise on signals as their amplitude decreases \cite{cogent}. A dashed curve marks the -1$\sigma$ boundary of the surface event distribution. A dotted line indicates the rise-time cut imposed to select bulk events (see text). }
\end{figure}

Dedicated background measurements were only possible on the day of PPC installation (10/19/2019). A large NaI[Tl] scintillator was used to study ambient gammas, confirming the presence of the dominant permanent $^{60}$Co component, in addition to a continuum of capture gammas from the primary containment wall, rapidly decreasing with energy, extending out to 11 MeV. The measured NaI[Tl] spectrum was unfolded into an isotropic gamma flux spanning eleven energy bins over the range 0.5-11.5 MeV, using a response matrix generated by MCNPX-PoliMi \cite{mcnpx}. MCNPX-PoliMi was employed for all other simulations in this work, using detailed geometries (Fig.\ 1). As a reference, this Rx-ON environmental flux  is 147 $\gamma$/cm$^{2}$s below 1.5 MeV, 20 $\gamma$/cm$^{2}$s for 1.5-3.5 MeV, 0.6 $\gamma$/cm$^{2}$s for 3.5-7 MeV, and 0.1 $\gamma$/cm$^{2}$s for 7-11 MeV.

The unfolded gamma flux was used to generate simulated  predictions in the  CE$\nu$NS ROI. The effect of photoneutrons generated by gammas above $\sim$7 MeV interacting in lead \cite{gnpb1,gnpb2} was found to be two orders of magnitude below present sensitivity. Special attention was paid to the possibility that coherent photon scattering from any energetic gammas able to reach the PPC might produce low-energy NRs able to compete with the CE$\nu$NS signal. This process has been proposed as a possible background affecting searches for low-mass dark matter candidates \cite{alan1,alan2}. Again, we find that its contribution to the ROI is more than three orders of magnitude lower than the present background level, which is dominated by elastic scattering from epithermal neutrons. This origin was confirmed by measurements employing a $^{3}$He counter in two configurations: bare, and surrounded by 6 cm of high-density polyethylene (HDPE) plus an external 0.6 mm layer of cadmium metal.

\begin{figure}[!htbp]
\includegraphics[width=.9 \linewidth]{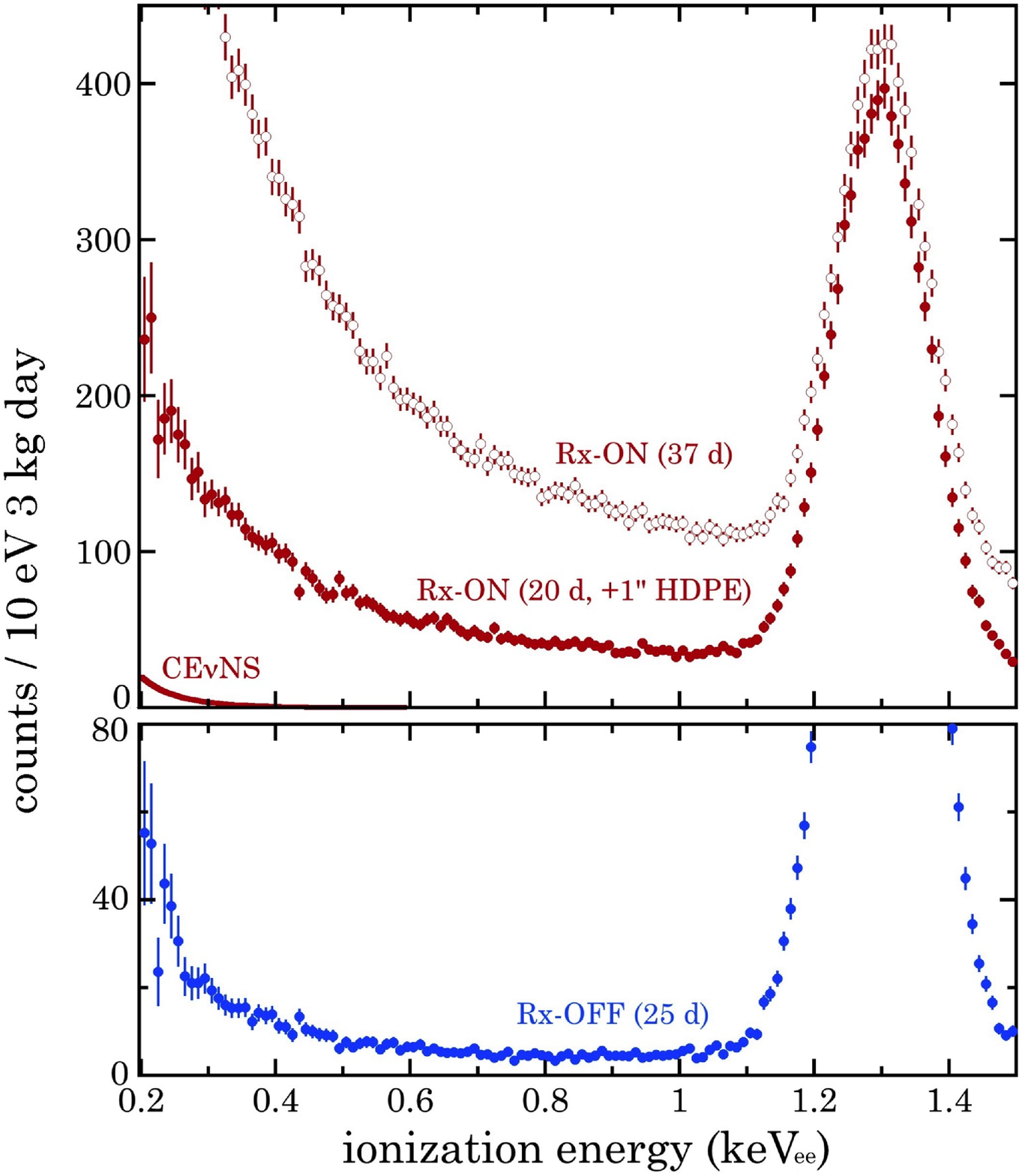}
\caption{\label{fig:epsart} Energy spectra for bulk PPC events passing all cuts during three initial runs, corrected for SA (see text). $^{71}$Ge activity  (1.29 keV$_{ee}$, \mbox{T$_{1/2}$=11.4 d}) has been drastically reduced in later runs. }
\end{figure}

Neutron fluence in the vicinity of a reactor is described by a three-component model: thermal, intermediate or epithermal, and fast \cite{iaea}. The epithermal spectrum follows a $E_{n}^{-(1+\alpha)}$ dependence on neutron energy $E_{n}$ over the range 0.55 eV (cadmium cut-off) to $\sim$1 MeV, with a value of $\alpha$ in excess of 0.2 for well-moderated cores \cite{oneovere1,oneovere2}. Our chosen detector and shielding materials have no strong scattering resonances or photoneutron reaction targets that could significantly distort this spectral shape. The lack of a significant fast component, expected from the several meters of moderator in the line-of-sight to core, was confirmed by the absence of any noticeable asymmetric ``shark tooth" peaks in the PPC spectrum. These are sensitive indicators of neutron inelastic scattering for $E_{n}\gtrsim$ 600 keV  \cite{shark1,shark2}. Using a MCNPX-PoliMi  response for both $^{3}$He configurations, the best-fit isotropic thermal and epithermal neutron fluxes for Rx-ON are 0.25 n/cm$^{2}$s and 0.57 n/cm$^{2}$s, respectively. For perspective, these correspond to an increase by a factor $\sim$100 over typical sea-level environmental values \cite{sea}.

Simulations of  PPC response to elastic scattering from the epithermal component display an excellent agreement with sub-keV Rx-ON data, in both spectral shape and rate. A best-match was obtained for $\alpha\!\simeq\!0.2$, indicating an intermediate neutron hardness slightly softer than that measured 17 m from the Brokdorf pressurized water reactor  \cite{conus2}. This presently-dominant background can be accurately described over the 0.2-1.0 keV$_{ee}$  CE$\nu$NS  ROI by a straightforward spectral shape model consisting of an energy-independent constant, plus an exponential component decreasing with increasing energy. Simplicity in a successful background model is to be expected, in view of the very narrow energy span of this ROI. However, as a precaution, this model was tested against its dependence on $\alpha$, on the choice of neutron cross-section MCNPX libraries for germanium, NR quenching factor model (i.e., fraction of NR energy expressed as ionization \cite{qf}, henceforth QF), threshold level of the inner veto, and modest addition of HDPE to shielding geometry. Its adequacy to satisfactorily describe
the simulated response to epithermal neutrons was confirmed in all cases. 

\begin{figure}[!htbp]
\includegraphics[width=.9 \linewidth]{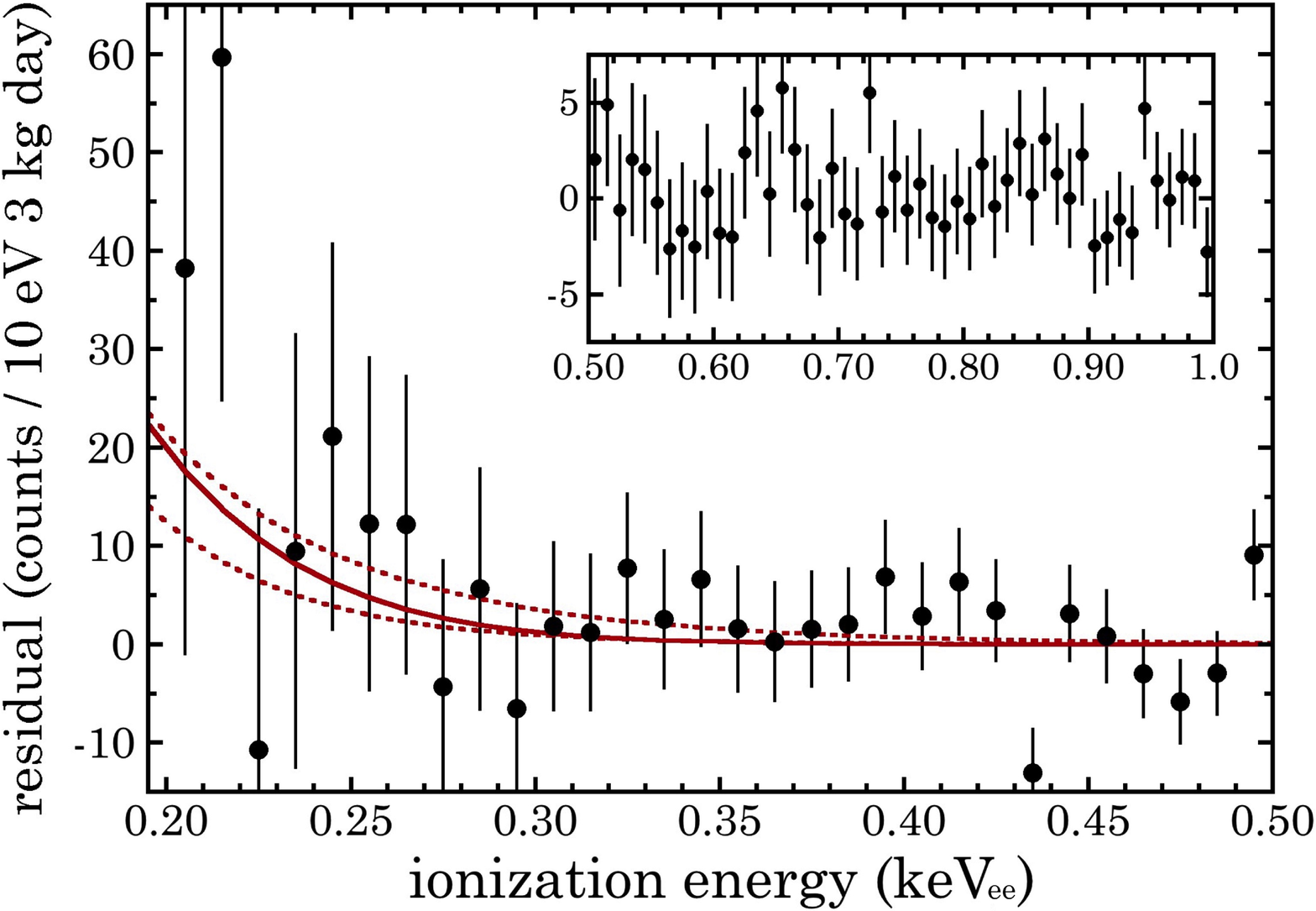}
\caption{\label{fig:epsart} Residual following subtraction of the best-fit epithermal neutron background in the alternative hypothesis model described in the text, for the 20 d Rx-ON dataset. The inset shows its continuation above 0.5 keV$_{ee}$. The expected CE$\nu$NS signal (solid line) involves a sub-keV QF recently measured in \cite{qf}, and a low-energy (anti)neutrino spectrum from \cite{Kopeikin}. This spectral choice produces CE$\nu$NS differential rates in good agreement with \cite{huber}. Dotted lines are for a constant QF of 30\% (top) and 20\% (bottom). These include the effect of energy resolution, adopting a Fano factor of 0.112 derived from the 10.37 keV$_{ee}$ $^{71}$Ge peak, in good agreement with \cite{fano}.   }
\end{figure}

Signal acceptance (SA) for CE$\nu$NS events was measured {\it in situ} using a programmable electronic pulser, set to provide a good replica of  fast rise-times from energy depositions in the bulk of the PPC, using as a reference the 1.29 keV$_{ee}$ peak from L-shell electron capture (EC) in $^{71}$Ge (Fig.\ 3), a byproduct of neutron capture in $^{70}$Ge. This peak and its K-shell  counterpart at 10.37 keV$_{ee}$ \cite{surface} were also used for energy calibration in the ROI. The choice of data cuts imposed was defined using these pulser runs and the first 48 h of Rx-ON operation, implementing a form of blind analysis. Additional data from a period of elevated ambient temperature and correspondingly-high FPGA triggering rate ($\times$5 that in present runs) were used to confirm that all near-threshold events originating in temperature-dependent detector noise fluctuations are removed by the strict quality cuts imposed (Fig.\ 3). 

A stringent condition accepting only preamplifier pulse rise-times shorter than 660 ns, was chosen to deliver 99\% rejection against subdominant surface events at the 0.2 keV$_{ee}$ analysis threshold (Fig.\ 4), while preserving a majority of $^{71}$Ge L-shell EC signals, and providing sufficient statistics of events passing cuts at the lowest energies. The prevalence of fast rise-time events at low energy, as is expected from elastic scattering in the bulk of the PPC by the dominant epithermal neutron background, is made noticeable by the grayscale of Fig.\ 4. 

Fig.\ 5 displays reconstructed (i.e., corrected for SA) low-energy spectra following all cuts. A conservative uncertainty of $\pm$5\% ($\pm$2\%) below (above) 300 eV$_{ee}$ is adopted for the SA. This accounts for the reproducibility observed in two separate pulser runs, their statistics, and  choice of function for their fit (Fig.\ 3).  Error bars in Fig.\ 5 combine the statistical error for events passing cuts, and this SA uncertainty. The spectra include the rejection of veto-coincident backgrounds, and account for a live-time correction from spurious veto coincidences, and from transient preamplifier saturation. Three spectra from initial runs are shown. The first corresponds to 25 days of data taken during a reactor refueling outage (Rx-OFF) spanning the period 10/28/2019 - 11/14/2019, in addition to a shorter technical outage during 12/28/2019 - 1/3/2020. This spectrum provides a reference for the best level of environmental background unrelated to reactor operation that can be expected in such a location, while using a compact shield. First Rx-ON runs (37 d) correspond to the period 11/17/2019-12/26/2019, excluding 3 days during which reactor power dropped below  100\%. Once the origin of the dominant epithermal neutron background was identified, a one inch-thick outermost layer of 5\% borated self-extinguishing HDPE was added to the shield. The thickness of moderator was limited to what could be hand-carried to the site, and installed during a brief two-hour visit on 3/6/2020. The observed reduction in ROI background by a factor of three during the ensuing 20 day run is in good agreement with simulated expectations, once again confirming epithermal neutrons as the dominant background.

\begin{figure}[!htbp]
\includegraphics[width=.9 \linewidth]{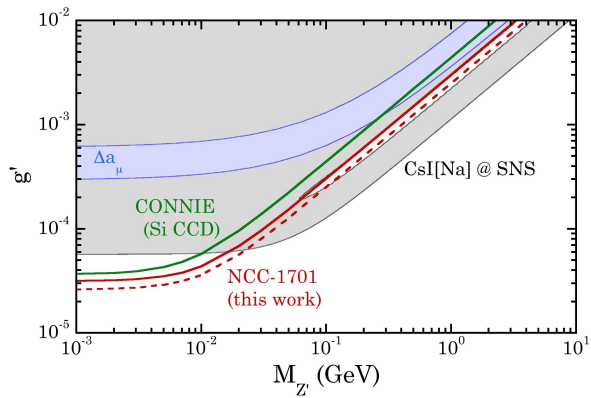}
\caption{\label{fig:epsart} 95\% C.L. limits on a universal flavor-conserving g\textsc{\char13} coupling from a neutral vector Z\textsc{\char13} boson, derived from present data (red line). A dashed  line shows the marginal improvement  from future PPC data able to constrain an excess CE$\nu$NS rate to 25\% above the SM expectation. Limits from CsI[Na] and silicon are also shown \cite{zcsi,zconnie}, together with a region (blue band) able to explain the discrepancy in the anomalous magnetic moment of the muon \cite{maxim,zcsi,zmuon}.}
\end{figure}

Fig.\ 5 also shows the expected CE$\nu$NS signal, illustrated for an energy-independent QF of 30\%. The best signal-to-background ratio at threshold from these initial runs is $\sim$1/10. For comparison, this was $\sim$1/4 during the first  detection of CE$\nu$NS from spallation neutrinos \cite{science,bjorn}. Neither exposure nor background level are sufficient to expect a positive CE$\nu$NS observation from  this initial dataset. However, our good understanding of the dominant background  invites to compare the performance of null and alternative hypotheses in fitting 0.2-1.0 keV$_{ee}$ ROI data. The null hypothesis contains three free parameters, describing the constant plus exponential terms in the epithermal neutron background model presented above. The alternative hypothesis adds to these terms a second exponential with fixed decay constant, and free amplitude. Its decay constant is chosen to provide an accurate approximation to the expected CE$\nu$NS signal shown in Fig.\ 6, derived using our most recent measurements of the sub-keV  QF in germanium \cite{qf} using $^{88}$Y/Be photoneutrons \cite{qfmyprl}. 

Fits to the 20 d Rx-ON spectrum were performed using a popular Markov Chain Monte Carlo (MCMC) ensemble sampler \cite{mcmc1,mcmc2}. Interestingly, the sampler rapidly converges to favor a non-zero CE$\nu$NS signal amplitude  of $37.5^{+26.5}_{-22.6}$ at 0.2 keV$_{ee}$ (units as in the vertical axis of Fig.\ 6, one-sigma uncertainty intervals). This is compatible within errors with the predicted CE$\nu$NS rate at this reactor location (Fig.\ 6). However, as expected from this short initial  exposure, the preference for the alternative hypothesis over the null, extracted from a likelihood ratio test, is presently very modest at p=0.27 ($\sim\!1.1 \sigma$ C.L.). Treated identically, a fit to the Rx-OFF dataset, for which no CE$\nu$NS signal can be expected, returns a smaller $16.8^{+10.3}_{-9.6}$ for this parameter, with similarly unremarkable preference (p=0.20, $\sim\!1.2 \sigma$ C.L.). This is compatible with the expected contribution  from $^{71}$Ge  M-shell EC \cite{supercdms} (0.158 keV$_{ee}$) above analysis threshold, for this run and present detector resolution. The abatement of this spectral component, the effect of alternative QF models \cite{qf} on the expected CE$\nu$NS signal, and a refined calculation of the antineutrino flux taking into consideration BWR power shape and core geometry will be treated within the context of upcoming datasets with higher exposure, lower background, and smaller uncertainties. 

As an example of the immediate utility of these initial datasets, we can exploit the 95\% upper confidence interval of the best-fit CE$\nu$NS signal amplitude mentioned above, to derive improved limits on a neutral vector boson Z\textsc{\char13} of mass M$_{Z'}$, able to mediate neutrino interactions in certain extensions of the SM \cite{davidc}. Based on  this interval, a maximum CE$\nu$NS rate of 4.3 times the SM prediction shown in Fig.\ 6 might be present in Rx-ON data. For comparison, this was 41 times the SM prediction for the previous Z\textsc{\char13} limit using low-noise CCDs at the Angra-II reactor \cite{zconnie}. Fig.\ 7 displays these new limits, derived following the same methodology as in \cite{zconnie}. The CCD limit in the figure is our own calculation, using a well-established QF model for silicon \cite{siqf1,siqf2}, resulting in an excellent agreement with the equivalent limit in \cite{zconnie}. We expect a similar improvement in sensitivity for other aspects of particle phenomenology accessible through reactor CE$\nu$NS  \cite{dimitrios,manfred}.

At the time of this writing, data-taking continues, following the installation of additional neutron shielding, with the goal of obtaining sufficient evidence for a first observation of reactor (anti)neutrino CE$\nu$NS. Profiting from the operational experience acquired with NCC-1701 at Dresden-II, we expect next-generation compact PPC assemblies to display a considerably lower background and threshold, and a larger signal acceptance. This may soon permit their operation as unattended reactor monitors, able to perform stably and continuously in harsh environments, providing a first instance of technological application for CE$\nu$NS.

This work was supported by awards DARPA W911NF1810222 and NSF PHY-1812702. We are deeply grateful to Exelon Corporation for their generosity in providing access to the Dresden-II reactor, and to \mbox{B. De Graaf} and D. Lim for their kind assistance, careful supervision, and frequent availability during reactor site interventions. We are thankful for the  roles played by L.R. Brennan, \mbox{T. Casagrande}, \mbox{C.M. Crane}, \mbox{D. Douglas}, S.U. Hansen, R.J. Herron, E.W. Hoppe, \mbox{D. Hornback}, \mbox{J. Jurewicz}, \mbox{W. Kouba}, \mbox{K.Y. Lee}, \mbox{H. Lippincott}, \mbox{J. Liptac}, S. Mell, J. Newby, \mbox{K. Petersen}, R. Rosner, J. Stevenson, V. Tang, \mbox{K. Taylor}, G. Ward, and M.C. Wrobel: their cumulative made this deployment possible. One of us (JIC) would like to thank his former COHERENT collaborators \mbox{P.S. Barbeau}, G.C. Rich, and K. Scholberg, for the impetus they provided to this initiative.

\bibliography{apssamp}

\end{document}